\documentclass[aps,prd,preprint,a4paper,showpacs,nofootinbib,superscriptaddress]{revtex4-2}
\usepackage{bbm}
\usepackage{amsmath}
\usepackage{multirow}
\usepackage{easybmat}
\usepackage{graphicx}
\usepackage{float}
\usepackage{amssymb}
\usepackage{mathrsfs}
\usepackage{subfigure}
\usepackage{amssymb}
\usepackage{hyperref}
\usepackage{epstopdf}

\usepackage[utf8]{inputenc}
\hypersetup{
    colorlinks=true,
    linkcolor=red,
    citecolor=blue,
}
\usepackage{color}
\usepackage[T1]{fontenc}
\usepackage{txfonts}

\begin{document}
 \newcommand{\bq}{\begin{equation}}
 \newcommand{\eq}{\end{equation}}
 \newcommand{\bqn}{\begin{eqnarray}}
 \newcommand{\eqn}{\end{eqnarray}}
 \newcommand{\nb}{\nonumber}
 \newcommand{\lb}{\label}
 
\title{Cuspy and fractured black hole shadows in a toy model with axisymmetry}

\author{Wei-Liang Qian}
\email{Email address: wlqian@usp.br}
\affiliation{Center for Gravitation and Cosmology, College of Physical Science and Technology, Yangzhou University, Yangzhou 225009, China}
\affiliation{Escola de Engenharia de Lorena, Universidade de S\~ao Paulo, 12602-810, Lorena, SP, Brazil}
\affiliation{Institute for theoretical physics and cosmology, Zhejiang University of Technology, Hangzhou, 310032, China}

\author{Songbai Chen}
\affiliation{Institute of Physics and Department of Physics, Hunan Normal University, 410081, Changsha, Hunan, China}

\author{Cheng-Gang Shao}
\affiliation{MOE Key Laboratory of Fundamental Physical Quantities Measurement, Hubei Key Laboratory of Gravitation and Quantum Physics, PGMF, and School of Physics, Huazhong University of Science and Technology, 430074, Wuhan, Hubei, China}

\author{Bin Wang}
\affiliation{Center for Gravitation and Cosmology, College of Physical Science and Technology, Yangzhou University, Yangzhou 225009, China}
\affiliation{School of Aeronautics and Astronautics, Shanghai Jiao Tong University, Shanghai 200240, China}

\author{Rui-Hong Yue}
\affiliation{Center for Gravitation and Cosmology, College of Physical Science and Technology, Yangzhou University, Yangzhou 225009, China}

\date{Aug. 2nd, 2021}

\begin{abstract}
Cuspy shadow was first reported for hairy rotating black holes, whose metrics deviate significantly from the Kerr one.
The non-smooth edge of the shadow is attributed to a transition between different branches of unstable but bounded orbits, known as the fundamental photon orbits, which end up at the light rings. 
In searching for a minimal theoretical setup to reproduce such a salient feature, in this work, we devise a toy model with axisymmetry, a slowly rotating Kerr black hole enveloped by a thin slowly rotating dark matter shell.
Despite its simplicity, we show rich structures regarding fundamental photon orbits explicitly in such a system. 
We observe two disconnected branches of unstable spherical photon orbits, and the jump between them gives rise to a pair of cusps in the resultant black hole shadow.
Besides the cuspy shadow, we explore other intriguing phenomena when the Maxwell construction cannot be established. 
We find that it is possible to have an incomplete arc of Einstein rings and a ``fractured'' shadow.
The potential astrophysical significance of the corresponding findings is addressed.

\end{abstract}

\maketitle
\newpage

\section{Introduction} \lb{section1}

The bending of light rays owing to the spacetime curvature constitutes one of the most influential predictions of General Relativity.
At its extreme form, the shadow casted by a black hole~\cite{agr-shadow-01, book-blackhole-DeWitt, agr-shadow-02, agr-shadow-03, agr-shadow-04, agr-shadow-06} is widely considered as an essential observable in the electromagnetic channel.
As the boundary of a black hole shadow is determined by the critical gravitational lensing of the radiation from nearby celestial bodies, it bears crucial information on spacetime geometry around the black hole.
With the prospect to directly probe the underlying theory of gravity in the strong-field region, the related topic has aroused much renewed curiosity in the past decade~\cite{agr-strong-lensing-shadow-52, agr-strong-lensing-shadow-08, agr-strong-lensing-shadow-09, agr-strong-lensing-shadow-55, agr-strong-lensing-21, agr-strong-lensing-shadow-54, agr-shadow-14, agr-shadow-15, agr-shadow-16, agr-shadow-17, agr-shadow-19, agr-shadow-20, agr-shadow-21, agr-shadow-23, agr-shadow-24, agr-shadow-25, agr-shadow-29, agr-shadow-26, agr-shadow-27, agr-strong-lensing-shadow-36} (for a concise review of the topic, see~\cite{agr-shadow-review-01}).
In particular, the supermassive black hole at the center of the M87 galaxy is being targeted by the Event Horizon Telescope (EHT)~\cite{agr-shadow-EHT-L01, agr-shadow-EHT-L04, agr-shadow-EHT-L05, agr-shadow-EHT-L06}.
Moreover, the developments regarding how to extrat information on the black hole in question from its silhouette open up a new avenue with promising possibilities~\cite{agr-strong-lensing-shadow-38, agr-strong-lensing-shadow-39, agr-strong-lensing-shadow-40, agr-shadow-EHT-05, agr-shadow-EHT-06, agr-shadow-EHT-07,agr-strong-lensing-EHT-08}.

The black hole shadow is defined by the set of directions in the observer's local sky where the ingoing null geodesics are originated from the event horizon.
In other words, no radiation is received by the observer at a certain solid angle due to the presence of the black hole.
Intuitively, the shape of the black hole shadow can be derived by analyzing the lower bound of the free-fall orbits that circulating the black hole in a compact spatial region.
Such a bound is closely associated with a specific type of null geodesics, dubbed fundamental photon orbits (FPOs)~\cite{agr-shadow-19, agr-shadow-04}, first proposed by Cunha {\it et al.}.
One may, by and large, argue that the edge of the shadow is furnished by the collection of the light rays that barely skim the unstable FPOs.
This is because a null geodesic that slightly deviates from an unstable FPO might marginally escape to the spatial infinity after orbiting the black hole for a multitude of times.
When tracing back in time, it is either originated from the black hole horizon or emanated by some celestial light source.
While the former constitutes part of the shadow by definition, a light ray associated with the latter, on the other hand, contributes to the image of the relevant celestial body.
In practical calculations, background radiation sources are placed on a sphere, referred to as the celestial sphere~\cite{book-blackhole-DeWitt}, (infinitely) further away from both the observer and the black hole.
Owing to the significant gravitational lensing, an infinite number of (chaotic) images of the entire celestial sphere pile up in the vicinity of the shadow edge~\cite{agr-shadow-10, agr-shadow-11, agr-shadow-12, agr-shadow-13}.

In the case of the Schwarzschild black hole, the relevant FPOs are the light rings (LRs), forming a photon sphere.
While for the Kerr one, the role of the FPO is carried by the spherical orbits~\cite{agr-shadow-07, agr-strong-lensing-22, agr-strong-lensing-23}.
The LRs are circular planar null geodesics, which by definition, is a particular type of FPOs associated with the axisymmetry of the relevant spacetimes.
In particular, it is understood that unstable LRs play a pivotal role in the strong gravitational lensing as well as shadow formation~\cite{agr-shadow-11, agr-shadow-19, agr-geodesic-03, agr-shadow-04, agr-geodesic-05}.
Stable LRs, being rather contrary to the nomenclature, might leads to the accumulation of different modes when the spacetime is perturbed~\cite{agr-geodesic-instability-02}.
Such a system is subsequently prone to nonlinear instabilities~\cite{agr-geodesic-instability-01}.
In the case of the Kerr black hole, the two LR solutions, restricted to the equatorial plane, are both unstable.
From the observer's viewpoint, on the shadow edge, they mark the two endpoints in the longitudinal direction.
The analyses of the black hole shadow in Kerr spacetime are simplified by the fact that the corresponding FPOs are of constant radius.
This is because the geodesic is Liouville integrable and separates in the Boyer-Lindquist coordinates~\cite{agr-bh-Kerr-03}.

In a generically stationary and axisymmetric spacetime, however, the separation of variables is often not feasible for the geodesic motion by choosing a specific coordinate chart.
As a result, the FPOs become more complicated and have to be evaluated numerically.
Nonetheless, it was pointed out~\cite{agr-shadow-19} the stability of LRs can be studied by employing the Poincar\'e maps.
Recently, Kerr black hole metrics with Proca hair were investigated, and a quantitatively novel shadow with cuspy edge was spotted~\cite{agr-shadow-19}.
Instead of a smooth shadow, the black hole silhouette is characterized by a pair of cusps at the boundary. 
The authors attributed the above feature to the sophisticated FPO structure, and in particular, to an interplay between stable and unstable FPOs.
To be specific, when compared with the case of Kerr metric, an additional stable branch of FPOs appears, which attaches both of its ends to that of two unstable branches of FPOs.
Consequently, a point of the cusp corresponds to a sudden transition between two FPOs from those unstable branches.
More lately, a similar characteristic was also reported~\cite{agr-shadow-20} in rotating non-Kerr black holes~\cite{agr-bh-Kerr-13}.
In quantum-gravity inspired models of regular black holes, cuspy shadow, dubbed ``dent-like'', was also observed in asymptotically safe gravity~\cite{agr-shadow-30}. 
For these cases, the metrics involved are rather complicated and mostly possess stable FPOs.
Apart from the above intriguing results, it still seems not very clear what is a minimal theoretical setup to reproduce a cuspy shadow edge.

If instead of vacuum, the black hole is surrounded by an accretion disk, a trespassing photon is likely subjected to inelastic scatterings.
As a result, it will deviate from its geodesic or even be entirely absorbed by an opaque disk.
However, if the disk is composed purely of dark matter, it is transparent to the photon.
This is because no observational signature regarding the interaction between the photon and dark matter particles has yet turned up, in any experiment designated to direct dark matter detection.
Nonetheless, the gravitational effect of the dark matter may still impact the null geodesics, and subsequently, the resultant black hole shadow.
For a spherical galactic black hole surrounded by a thick dark matter cloud, it was argued that observable deviation from its Schwarzschild counterpart might be expected~\cite{agr-shadow-29}.  
More lately, in studying the rotating dirty black holes~\cite{agr-shadow-26}, the authors found that although the existence of the dark matter modifies the size of the shadow, the D-shaped contour almost remains unchanged.

Moreover, although the physical nature remains largely elusive, in literature, many interesting substructures in the dark matter halos and sub-halos have been speculated~\cite{agr-dark-matter-02, agr-axion-dark-matter-04, agr-dark-matter-06, agr-dark-matter-07, agr-dark-matter-condensate-08}.
Among a large variety of alternatives, the venerable $\Lambda$CDM model indicates that a discontinuity in the matter distribution might be triggered by the presence of the dark matter~\cite{agr-dark-matter-06, agr-dark-matter-07}.
In this regard, the exploration of the rich substructure regarding the dark matter halo is closely related to our understanding of the underlying physics.
Indeed, the resultant substructures predicted by the theoretical models may, in turn, serve to discriminate between different interpretations about the nature of dark matter.
Nonetheless, it is not clear whether a discontinuity in the matter distribution may further appreciably distort the black hole shadow, particularly to the extent such modification becomes potentially observable.

In fact, recently, it was shown analytically~\cite{agr-qnm-lq-01} that a discontinuity in the effective potential significantly affects the asymptotic properties of quasinormal modes.
As the last phase of a merger process, such a dramatic change in the quasinormal ringing may potentially lead to observable effects.
In fact, discontinuity is present at the surface of compact celestial objects, and the numerical calculations of the curvature modes have indeed confirmed such a nontrivial consequence~\cite{agr-qnm-star-07}.
Furthermore, it was recently speculated that discontinuity due to a thin disk of matter provides an alternative mechanism of the echo phenomenon~\cite{agr-qnm-echoes-20}. 
Considering that the quasinormal frequencies at the eikonal limit are closely connected with the shadow and photon sphere in spherical metrics~\cite{agr-geodesic-06, agr-shadow-24}, 
it is natural to ask whether some discontinuity out of the horizon of the black hole might lead to meaningful implications in the context of the black hole shadow.
This is the primary motivation of the present study.

The present study continues to pursue further discussions on black hole shadows concerning the role of dark matter surrounding the black holes.
By simplifying the dark matter envelope to a thin shell wrapped around the black hole, the mass distribution is concentrated in an infinitesimal layer so that we have a sharp discontinuity in the effective potential.
We will show that such stationary axisymmetric configuration is physically plausible as the Israel-Lanczos-Sen junction condition is satisfied at the slow rotation limit.
The resultant spacetime possesses two branches of unstable FPOs but not any stable FPO. 
Our analysis reveals a sudden jump between the different branches, which results in a cusp on the boundary of the black hole shadow.
Moreover, it is argued that the transition point can be determined using the Maxwell construction, which is reminiscent of the Gibbs conditions for the phase transition.
We also investigate other intriguing possibilities regarding different model parameterizations, which include the cases involving an incomplete arc of the Einstein rings and fractured shadow edge.
The astrophysical significance of the present findings is addressed.

The remainder of the manuscript is organized as follows.
In the next section, we present our model and the mathematical framework for evaluating the null geodesics as well as the associated celestial coordinates.
We study the properties of the FPOs and discuss the relevant criterion to determine the black hole shadow.
In Sec.~\ref{section3}, for a specific choice of the metric, we show that the Maxwell construction can be utilized to locate the transition point on the shadow edge, where the cusp is present.
The discussions also extend to other meaningful metric parameterizations, where one elaborates on two additional intriguing scenarios.
Further discussions and concluding remarks are given in the last section.
We relegate the mathematical derivations regarding the Israel-Lanczos-Sen junction condition to the appendix.

\section{A dark matter shell toy model} \lb{section2}

In this section, we first present the proposed toy model and then proceed to discuss the FPOs of the relevant metric as well as their connection with the black hole shadow.
For the purpose of the present study, we consider a stationary axisymmetric metric with the following form in the Boyer-Lindquist coordinates $(t, r, \theta, \varphi)$
\begin{eqnarray}
ds^2=&-&\left(1-\frac{2Mr}{\Sigma}\right)dt^2-\frac{4Mar\sin^2\theta}{\Sigma}dtd\varphi+\frac{\Sigma}{\Delta}dr^2+\Sigma d\theta^2 \nonumber\\
&+&\left(r^2+a^2+\frac{2Ma^2r\sin^2\theta}{\Sigma}\right)\sin^2\theta d\varphi^2 ,
\label{metric_Kerr}
\end{eqnarray}
where
\begin{eqnarray}
\Delta &=& r^2-2Mr+a^2, \nonumber \\
\Sigma &=& r^2+a^2\cos^2\theta , \nonumber
\end{eqnarray}
and
\bqn
M &=&
\left\{\begin{array}{cc}
M_{\mathrm{BH}}    &  r\le r_{\mathrm{sh}}  \cr\\
M_{\mathrm{TOT}} &  r_{\mathrm{sh}} < r <\infty 
\end{array}\right. , \nonumber \\
a &=&
\left\{\begin{array}{cc}
a_{\mathrm{BH}}    &  r\le r_{\mathrm{sh}}  \cr\\
a_{\mathrm{TOT}} &  r_{\mathrm{sh}} < r <\infty 
\end{array}\right. ,
\lb{Ma_cut}
\eqn
where $M_{\mathrm{BH}}, a_{\mathrm{BH}}, M_{\mathrm{TOT}}$ and $a_{\mathrm{TOT}}$ are constants, the rotation parameters $|a|\ll 1$, $r_{\mathrm{sh}}$ is the location of thin layer of dark matter.
For both the regions $r\le r_{\mathrm{sh}}$ and $r_{\mathrm{sh}} < r <\infty$, the metric coincides with that of a Kerr one, satisfying the Einstein's equation in vaccum.

Inside the dark matter envelop, the above spacetime metric describes a Kerr black hole with the mass $M=M_{\mathrm{BH}}$ and angular momentum $J=a_{\mathrm{BH}}M_{\mathrm{BH}}$ sitting at the center.
For an observer sitting far away ($r\gg r_{\mathrm{sh}}$), they are essentially dealing with a rotating spacetime with the mass $M=M_{\mathrm{TOT}}$ and angular momentum per unit mass $a=a_{\mathrm{TOT}}$.
The discontinuity at $r=r_{\mathrm{sh}}$ indicates a rotating (infinitesimally) thin shell of mass wrapping around the central black hole.
It is important to note that for the above metric to be physically meaningful, it must be validated against the Israel-Lanczos-Sen's junction conditions~\cite{agr-collapse-thin-shell-03}.
In particular, the induced metrics onto the shell from both interior and exterior spacetimes must be isometric.
While relegating the details to the appendix, we argue that at the slow rotation limit, the first junction condition can be fulfilled.
Therefore, we will only consider the choice of parameters satisfying $\left|a_{\mathrm{BH}}\right|, \left|a_{\mathrm{TOT}}\right| \ll 1$.

The null geodesic of a photon in a pure Kerr spacetime satisfying the following system of equations~\cite{agr-shadow-07}
\begin{eqnarray}
\Sigma\frac{dt}{d\lambda} &=& \frac{(a^2+r^2)((a^2+r^2)E-aL)}{\Delta}+aL-a^2 E\sin^2\theta , \nonumber \\
\Sigma\frac{d\varphi}{d\lambda} &=& \frac{a((a^2+r^2)E-aL)}{\Delta}+L\csc^2\theta - aE ,\nonumber \\
\Sigma\frac{dr}{d\lambda} &=& \pm\sqrt{R} ,\nonumber \\
\Sigma\frac{d\theta}{d\lambda} &=& \pm\sqrt{\Theta} ,
\lb{Kerr_geodesic}
\end{eqnarray}
where 
\begin{eqnarray}
R &=& -\Delta(Q+(L - aE)^2) + ((a^2+r^2)E -aL)^2 , \nonumber \\
\Theta &=& Q+\cos^2\theta\left(a^2E^2-\frac{L^2}{\sin^2\theta}\right) .\nonumber 
\end{eqnarray}
Here $E$, $L$, and $Q$ are the energy, angular momentum, and the Carter constant of the photon.
For our present case, the analysis of the null-geodesic motion can be achieved by implementing a simple modification.
For light rays propagating inside a given region, namely, $r<r_{\mathrm{sh}}$ or $r>r_{\mathrm{sh}}$, its motion is governed by Eq.~\eqref{Kerr_geodesic} with the metric parameters given by Eq.~\eqref{Ma_cut}.
The difference occurs when the light ray crossing the dark matter layer at $r=r_{\mathrm{sh}}$.
For instance, let us consider a free photon that escapes from the inside of the thin shell.
Due to the singularity in the derivatives of the metric tensor at $r=r_{\mathrm{sh}}$, the photon's trajectory will suffer a deflection as it traverses the shell.
However, the values of $E$, $L$, and $Q$ remain unchanged during the process and thus can be utilized to unambiguously match the geodesics on both sides of the shell. 
This is because these constants of motion are derived by the corresponding Killing objects implied by the axisymmetry of the spacetime in question.
Subsequently, when given one point on the trajectory, a null-geodesic motion is entirely determined by a pair of values,
\begin{eqnarray}
\eta &=& \frac{L}{E} , \nonumber \\
\xi &=& \frac{Q}{E^2}.
\lb{etaxi}
\end{eqnarray}
This can be easily seen by rescaling the affine parameter $\lambda \to \lambda'=\lambda E$ in Eq.~\eqref{Kerr_geodesic}.
Moreover, due to the axisymmetry, the separation of variables is still feasible for the present case, and therefore, all the FPOs are spherical orbits as for the Kerr metric.
The spherical orbit solution can be obtained by analyzing the effective potential associated with the radial motion, which separates from those of angular degrees of freedom.
To be specific, the third line of Eq.~\eqref{Kerr_geodesic} can be rewritten as
\begin{eqnarray}
\Sigma^2\dot{r}^2 + V_{\mathrm{eff}}=0 ,
\lb{Kerr_radial_motion}
\end{eqnarray}
where the effective potential $V_{\mathrm{eff}}$ reads
\begin{eqnarray}
\frac{V_{\mathrm{eff}}}{E^2}=\left(a^2+r^2-a \xi\right)^2-\left(r^2+a^2-2r\right)\left(\eta+(\xi-a)^2\right) .
\lb{Kerr_radial_V}
\end{eqnarray}
Similar to the analysis of the planetary motion in Newtonian gravity, the spherical orbits are determined by the extremum of the effective potential, namely, $V_{\mathrm{eff}}=\partial_r V_{\mathrm{eff}}=0 $.
One finds
\begin{eqnarray}
\eta &=& \frac{r_0^3-3Mr_0^2+a^2r_0+Ma^2}{a(M-r_0)} , \nonumber \\
\xi &=& \frac{r_0^2(3r_0^2+a^2-\eta^2)}{r_0^2-a^2} ,
\lb{etaxi_geodesic}
\end{eqnarray}
where $r_0$ is the radius of the spherical orbit.
These orbits are unstable since the encountered extremum is a local maximum.
Moreover, the fact that all FPO are spherical orbits is related to the uniqueness of the above local maximum. 

On the other hand, for an observer located at (asymptotically flat) infinity with zenithal angle $\theta_0$, the boundary of the black hole is governed by those null geodesics that marginally reach them.
By assuming that the entire spacetime is flat, the ``visual" size of the black hole can be measured by slightly ``tilting their head'' (or in other words, by an infinitesimal displacement of their location).
To be specific, when projected onto the plane perpendicular to the line of sight, the size of the image in the equatorial plane and on the axis of symmetry can be obtained by the derivatives of the angular coordinates $(\varphi, \theta)$~\cite{book-blackhole-DeWitt}.
These derivatives can be calculated explicitly using the asymptotical form of the geodesic, namely, Eq.~\eqref{Kerr_geodesic} evaluated at the limit $r\to \infty$.
We have
\begin{eqnarray}
\alpha &=& \lim\limits_{r\to\infty} \left(\left.-r^2\sin\theta_0\frac{d\varphi}{dr}\right|_{\theta\to\theta_0}\right)=-\eta \csc\theta_0 , \nonumber \\
\beta &=& \lim\limits_{r\to\infty} \left(\left. r^2\frac{d\theta}{dr}\right|_{\theta\to\theta_0}\right)=\pm\sqrt{\xi+a_{\mathrm{TOT}}^2\cos^2\theta_0-\eta^2\cot^2\theta_0} ,
\lb{alphabeta}
\end{eqnarray}
where the pair $(\eta, \xi)$ are dictated by the geodesic of the photon in question.
The coordinates in terms of $\alpha$ and $\beta$ are often referred to as the celestial coordinates in the literature~\cite{book-blackhole-DeWitt}.
By collecting all coordinate pairs $(\alpha, \beta)$, one is capable of depicting the apparent silhouette of the black hole.

As discussed above, the relevant null geodesics that potentially contribute to the shadow edge are the FPOs.
In contrary to the evaluation of the celestial coordinates, which involves the asymptotic behavior of the metric, the FPOs are determined by spacetime properties in the vicinity of the horizon.
Although all the FPOs for our metric are spherical orbits, the presence of a rotating thin shell leads to some interesting implications.

In what follows, let us elaborate on the properties of the FPO and their connection with the black hole shadow.
First, consider a FPO solution for the pure Kerr spacetime with $M=M_{\mathrm{BH}}, a=a_{\mathrm{BH}}$.
It will also be qualified as an FPO for the metric defined in Eq.~\eqref{Ma_cut}, if and only if the radius of the corresponding spherical orbit $r_0$ satisfies $r_0 < r_{\mathrm{sh}}$.
Likewise, a FPO solution with $r_0 < r_{\mathrm{sh}}$ for the pure Kerr spacetime with $M=M_{\mathrm{TOT}}, a=a_{\mathrm{TOT}}$ does not exist physically for the metric under consideration. 

Secondly, we note that {\it not every} FPO contributes to the edge of the black hole shadow.
Let us consider, for instance, a photon moves along a spherical orbit right outside the shell with $r_0 = r_{\mathrm{sh}}+0^+$.
When its trajectory is perturbed and let us assume that the photon spirals slightly inward.
As the photon conserves its values of $(\eta, \xi)$, at the moment it intersects the infinitesimally thin shell, the trajectory is promptly deflected from the tangential direction perpendicular to the radius.
This implies that it no longer stays in the vicinity of any spherical orbits, namely, the FPO for the region $r_0 < r_{\mathrm{sh}}$.
Subsequently, the photon will spiral into the event horizon rather quickly instead of critically orbiting the black hole for an extensive number of times beforehand.
This, in turn, indicates the photon is mapped onto a pixel disconnected from those associated with the FPOs of the region $r_0 < r_{\mathrm{sh}}$, which constitute the shadow edge.
The above heuristic arguments can be reiterated in terms of the fact that the pair of values $(\eta, \xi)$ for an FPO of the outer region $r_0 > r_{\mathrm{sh}}$ does not, in general, corresponds to that of an FPO of the inner region $r_0 < r_{\mathrm{sh}}$.
Therefore, the photon which skims the thin layer of dark matter on the outside, by and large, does not contribute to the edge of the black hole shadow. 
Now, one may proceed to consider a peculiar case, where the pair of values $(\eta, \xi)$ for an FPO in the outer region matches that of an FPO in the inner region.
Therefore when the trajectory of the former is perturbed and the photon eventually traverses the thin shell, it will still stay in the vicinity of the latter and eventually contributes to the edge of the shadow.
Since the values of $(\eta, \xi)$ for both FPO are the same, according to Eq.~\eqref{alphabeta}, they also contribute to the same pixel in the celestial coordinates.
This is precisely the Maxwell condition that we will explore further in the next section.
It is worth noting that, even if an FPO does not directly contribute to the shadow edge, it is still subjected to strong gravitational lensing and therefore possibly leads to a nontrivial effect.

Moreover, we note that the inverse of the above statement is still valid.
In other words, the edge of the black hole shadow is entirely furnished by the FPOs in either region of the spacetime.
If some FPOs in the outer region $r_0 > r_{\mathrm{sh}}$ contributes to the shadow edge, the section of the shadow boundary is identical to those of a Kerr black hole with $M=M_{\mathrm{TOT}}, a=a_{\mathrm{TOT}}$.
However, if some FPOs in the inner region $r_0 < r_{\mathrm{sh}}$ contributes to the shadow edge, due to Eq.~\eqref{alphabeta}, the corresponding section of the black hole silhouette is different from that of the Kerr black hole that sits inside the thin shell.

Before proceeding further, we summarize the key features regarding the FPOs in the present model and their connection with the black hole shadow edge as follows
\begin{itemize}
  \item The null-geodesic motion is determined by a pair of conserved quantity $(\eta, \xi)$.
  \item The black hole shadow is a projection of asymptotic light rays onto a plane perpendicular to the observer's line of sight, and any point on its edge is governed by the two-dimensional orthogonal (celestial) coordinates consisting of $(\alpha, \beta)$.
  \item The boundary of the black hole shadow is largely determined by the unstable FPOs\footnote{For the particular case where there is no horizon, the shadow edge might not be entirely furnished by FPOs, as discussed in Refs.~\cite{agr-strong-lensing-shadow-08, agr-strong-lensing-shadow-09}.}, but some FPO may not contribute to the shadow edge.
  \item Due to the presence of the thin shell, some formal FPO solutions for the pure Kerr spacetime are not physically relevant.
  \item When the values of $(\eta, \xi)$ of a particular FPO on one branch match those of another FPO on a different branch, both FPOs contribute to the same point in the celestial coordinates, probably on the shadow edge.
\end{itemize}

\section{The Maxwell construction and black hole shadow} \lb{section3}

In the last section, we discuss the close connection between the unstable FPOs and the black hole shadow edge.
It is pointed out that the null-geodesic motion can be determined in terms of the pair of values $(\eta, \xi)$.
As this is the same number of degrees of freedom to locate a specific point on the celestial coordinates, the dual $(\eta, \xi)$ of an FPO can be used to map onto the corresponding point on the shadow edge in the celestial coordinates.
To be specific, the transition point on the shadow edge can be identified by matching $(\eta, \xi)$ for two FPOs from different branches, namely,
\begin{eqnarray}
\eta(r^{\mathrm{cusp}}_{\mathrm{BH}}) &=& \eta(r^{\mathrm{cusp}}_{\mathrm{TOT}}) , \nonumber \\
\xi(r^{\mathrm{cusp}}_{\mathrm{BH}}) &=& \xi(r^{\mathrm{cusp}}_{\mathrm{TOT}}) ,
\lb{Maxwell_01}
\end{eqnarray}
where $r^{\mathrm{cusp}}_{\mathrm{BH}} < r^{\mathrm{cusp}}_{\mathrm{TOT}}$ are two distinct FPO solutions, belong to the two distinct unstable branches of FPOs.
Since the established condition is between two sets of quantities, it is reminiscent of the Maxwell construction (e.g. in terms of the chemical potentials) in a two-component system~\cite{phase-01,phase-02}.
On a rather different ground, such a construction was derived from the Gibbs conditions for the phase transition in a thermodynamic system.
For the present context, the pair $(\eta, \xi)$ determined by Eq.~\eqref{Maxwell_01} is mapped to $(\alpha, \beta)$ in the celestial coordinates, which subsequently gives rise to a cusp on the shadow edge.
Such a salient feature is similar to what has been discovered earlier~\cite{agr-shadow-19, agr-shadow-20} using more sophisticated black hole metrics.

The present section is devoted to investigating different scenarios emerging from the proposed model.
We show that, due to an interplay between different branches of unstable FPOs and the location of the discontinuity introduced by the thin shell, the resultant black hole shadow presents a rich structure.
The following discussions will be primarily concentrated on three sets of model parameters, given in Tab.~\ref{tb_shadow}.
The choice of the parameters aims at enumerating all relevant features in the present model, in terms of the feasibility of the Maxwell construction, as well as the different roles carried by the FPO.
In particular, in the first case, the Maxwell construction can be established.
Besides, the parameters are chosen so that unstable FPOs contribute both to black hole shadow edge and {\it metastable} states, after the unphysical ones are excluded.
In the other two cases, on the other hand, one cannot find such a transition between different branches of FPOs via the Maxwell construction.
However, two physically interesting scenarios are observed for these cases.
The second set of parameters leads to an incomplete section of Einstein rings, while the third set gives rise to a fractured black hole shadow.
For simplicity, for all three cases, we set $M_{\mathrm{BH}}=1.0$, while satisfying $\left|a_{\mathrm{BH}}\right|, \left|a_{\mathrm{TOT}}\right| \ll 1$.
By using these parameters, the four rightmost columns list the calculated radii of LRs.
The latter correspond to the radial bounds for the spherical orbits if there were no constraints associated with the thin shell.
We note if one employs smaller values for $\left|a_{\mathrm{BH}}\right|$ and $\left|a_{\mathrm{TOT}}\right|$, all the observed features remain.

\begin{table}[htb]
\begin{center}
{\begin{tabular}{ c||c| c| c| c| c||c| c| c| c }
\hline
set & $M_{\mathrm{BH}}$ & $a_{\mathrm{BH}}$ & $M_{\mathrm{TOT}}$ & $a_{\mathrm{TOT}}$ &  $r_{\mathrm{sh}}$ & $r^-_{\mathrm{BH}}$ & $r^+_{\mathrm{BH}}$ & $r^-_{\mathrm{TOT}}$ & $r^+_{\mathrm{TOT}}$ \\ \hline
$1$ &  $1.0 $ & $0.2$ & $1.01$ & $0.01$ & $3.028$ & $2.759$ & $3.223$ & $3.018$ & $3.042$ \\
$2$ &  $1.0 $ & $0.2$ & $1.10$ & $0.01$ & $3.310$ & $2.759$ & $3.223$ & $3.288$ & $3.312$ \\
$3$ &  $1.0 $ & $0.1$ & $1.05$ & $0.20$ & $3.000$ & $2.882$ & $3.113$ & $2.910$ & $3.373$  \\ \hline
\end{tabular}}
\end{center}
\caption{The three sets of paramters utilized in the present study.
The four right most columns give the calculated radii of the LRs, $r^-_{\mathrm{BH}}, r^+_{\mathrm{BH}}$ and $r^-_{\mathrm{TOT}}, r^+_{\mathrm{TOT}}$, 
for the regions $r<r_{\mathrm{sh}}$ and $r>r_{\mathrm{sh}}$ of the metric, respectively. 
}
\label{tb_shadow}
\end{table}

\begin{figure}
\begin{tabular}{cc}
\vspace{0pt}
\begin{minipage}{225pt}
\centerline{\includegraphics[width=200pt]{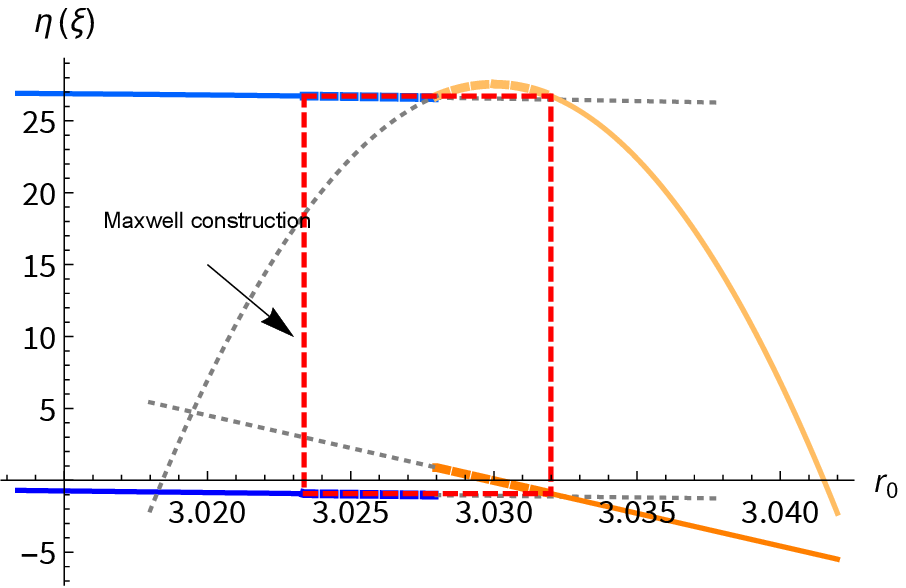}}
\end{minipage}
&
\begin{minipage}{225pt}
\centerline{\includegraphics[width=200pt]{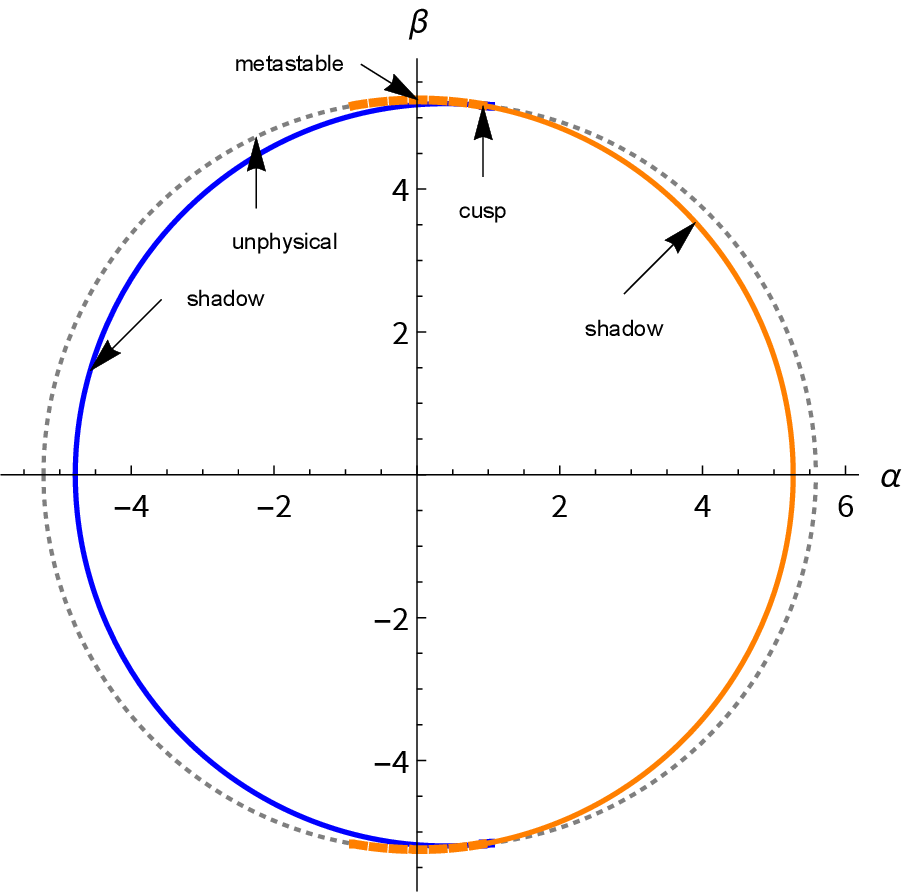}}
\end{minipage}
\end{tabular}
\centerline{\includegraphics[width=200pt]{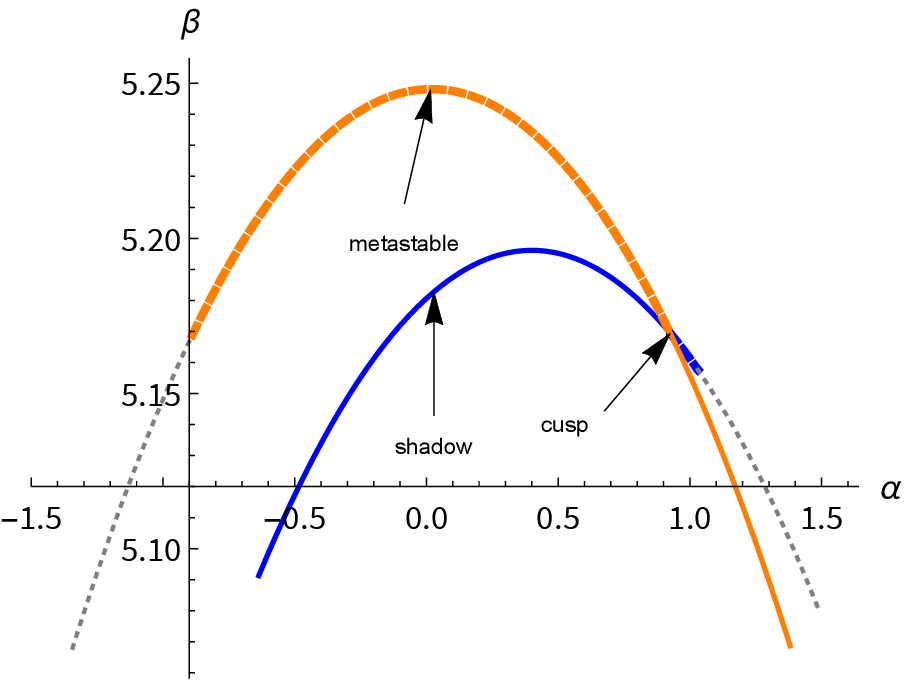}}
\renewcommand{\figurename}{Fig.}
\caption{(Color online)
The Maxwell construction and the corresponding black hole shadow with cusp.
The blue curves denote the FPOs associated with the metric for the region $r<r_{\mathrm{sh}}$, while the orange ones are those for the region $r> r_{\mathrm{sh}}$. 
The solid blue and orange curves (labeled ``shadow'') are the collections of FPOs that contribute to the shadow edge.
The dashed blue (barely visible) and orange curves (labeled ``metastable'') represent those FPOs that do not directly give rise to the shadow edge.
The dotted gray curves (labeled ``unphysical'') correspond to the FPO solutions that are not physically permitted.
Top left: The Maxwell construction, shown in the dashed red rectangle, establishes the transition point between the two branches of unstable FPOs in terms of $\eta$ and $\xi$ as functions of orbit radius $r_0$.
The curves with dark colors (dark blue and dark orange) are for $\eta=\eta(r_0)$, while those with light colors (light blue and light orange) are for $\xi=\xi(r_0)$.
The unstable FPOs excluded from the shadow edge by the Maxwell construction are denoted as ``metastable'' due to their resemblance to the metastable states in a thermodynamical system.
Top right: The corresponding shadow edge is shown in solid blue and orange curves, where the transition point is labeled by ``cusp''.
The eyelash shape extension of the shadow edge, shown in dashed curves, may still lead to a strong gravitational lensing effect.
Bottom: The same as the top right plot, where the region in the vicinity of the ``cusp'' is amplified.}
\label{fig1_cusp}
\end{figure}

We first consider the first set of model parameters given in Tab.~\ref{tb_shadow}, and the calculated cuspy black hole shadow is shown in Fig.~\ref{fig1_cusp}.
To give a more transparent presentation, for the figures, we adopt the following conventions. 
The FPOs associated with the edge of the black hole shadow are shown by solid curves.
Meanwhile, the FPOs that are valid null geodesics of the metric but irrelevant to the shadow are depicted in dashed curves.
The gray dotted curves are unphysical FPO solutions.
As discussed in the last section, they must be excluded due to the physical constraint related to the thin shell.
It is observed that the resultant spacetime is featured by two disconnected branches of unstable FPOs.
The FPOs associated with the inner region ($r<r_{\mathrm{sh}}$) are shown in solid and dashed blue curves.
Those associated with the outer region ($r>r_{\mathrm{sh}}$) are represented by solid and dashed orange curves. 
As shown in the left plot of Fig.~\ref{fig1_cusp}, the Maxwell construction, Eq.~\eqref{Maxwell_01}, is indicated by the dashed red rectangle.
It corresponds to the transition point (labeled ``cusp'') in the right plot.
However, it is worth noting that, different from previous studies~\cite{agr-shadow-19, agr-shadow-20}, the present metric does not possess any stable FPO.
Therefore, the latter is not a necessary condition for the presence of the cusp.

The Maxwell construction give $r^{\mathrm{cusp}}_{\mathrm{BH}} \simeq 3.023  < r^{\mathrm{cusp}}_{\mathrm{TOT}} \simeq 3.032$.
Our choice of $r_{\mathrm{sh}}=3.028$ ensures that there is still some room for an interesting feature.
In Fig.~\ref{fig1_cusp}, the cusp divides both branches of unstable FPOs into two parts, shown in solid and dashed curves, while labeled ``shadow'' and ``metastable'' in the right plot, respectively.
The FPOs on one side of the cusp constitute the edge of the black hole shadow.
The FPOs on the other side are, though not contributing to the shadow, still subjected to strong gravitational lensing.
As a result, they demonstrate themselves as a particular lensing pattern connected to the cusp.
This is nothing but the ``eyelash'' feature discussed in Ref.~\cite{agr-shadow-19}.
They are labeled ``metastable'' due to their apparent resemblance to the metastable states in thermodynamics, associated with superheated and subcooled states.
In other words, such states are allowed physically but do not directly contribute to the shadow edge in question.  
Our particular choice of the metric parameters given in the first set of Tab.~\ref{tb_shadow}, namely, $r_{\mathrm{TOT}}^- < r_{\mathrm{sh}} < r^{\mathrm{cusp}}_{\mathrm{TOT}}$ and $r^{\mathrm{cusp}}_{\mathrm{BH}} < r_{\mathrm{sh}} < r_{\mathrm{BH}}^+$, implies that the ``eyelash'' is present for both branches after the removal of unphysical FPOs.
For instance, the dashed orange eyelash shown in the top left and bottom plots of Fig.~\ref{fig1_cusp} corresponds to the FPOs with their orbital radii $r^{\mathrm{cusp}}_{\mathrm{BH}} < r_0 < r_{\mathrm{sh}} $.

\begin{figure}
\begin{tabular}{cc}
\vspace{0pt}
\begin{minipage}{225pt}
\centerline{\includegraphics[width=200pt]{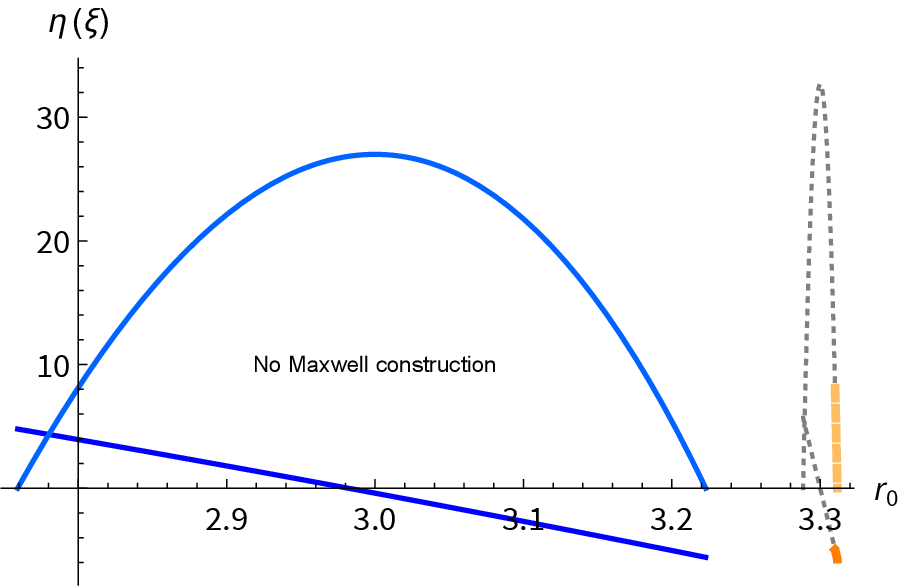}}
\end{minipage}
&
\begin{minipage}{225pt}
\centerline{\includegraphics[width=200pt]{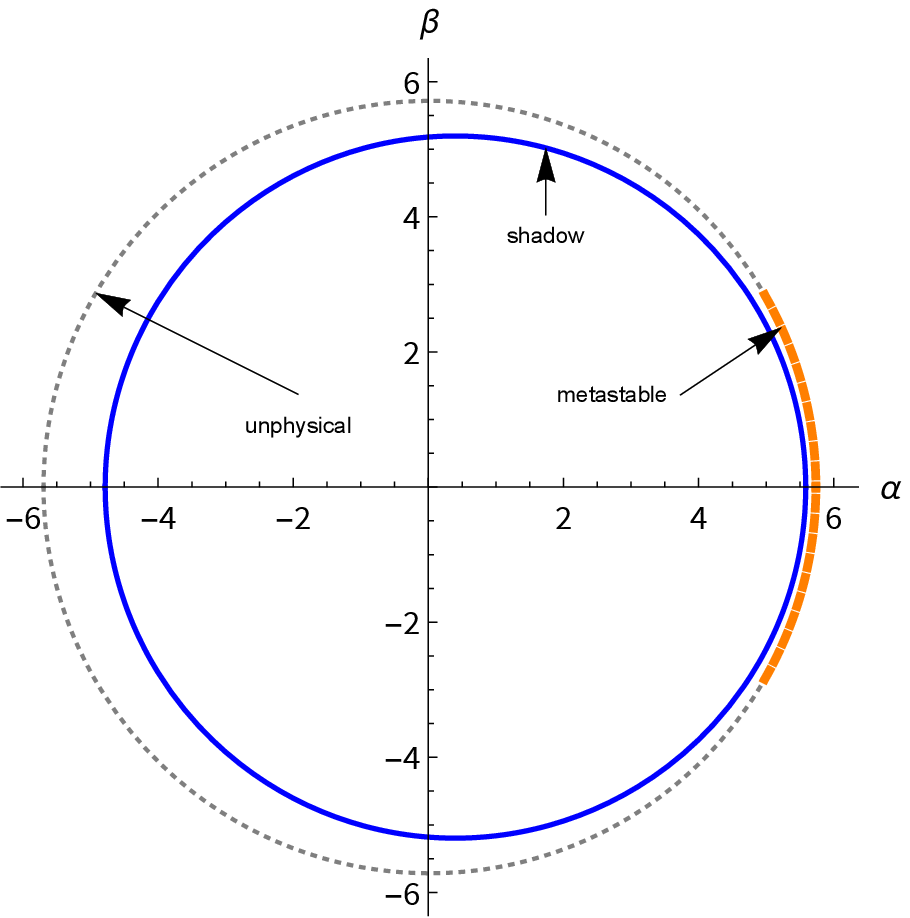}}
\end{minipage}
\end{tabular}
\renewcommand{\figurename}{Fig.}
\caption{(Color online)
The black hole shadow is surrounded by an incomplete section of Einstein rings.
We adopt the same convention for the line types and colors as Fig.~\ref{fig1_cusp}.
Left: No Maxwell construction can be established for this case.
Right: The resultant black hole shadow edge is shown by the closed solid curve.
The unstable FPOs, associated with the metric in the region $r> r_{\mathrm{sh}}$, form a section of arc, which might lead to an incomplete ring.
}
\label{fig2_half_ring}
\end{figure}

Now, we move to consider the other two scenarios where the Maxwell construction cannot be encountered.
In both cases, the resulting black hole shadow does not possess any cusp, but still, noticeable features are observed.
In Fig.~\ref{fig2_half_ring}, we present the results obtained for the second set of metric parameters given in Tab.~\ref{tb_shadow}.
From the left plot, the Maxwell construction can not be established, and the resultant boundary of the black hole shadow is subsequently determined by the metric of the Kerr black hole sitting at the center.
However, since $r_{\mathrm{sh}} < r^+_{\mathrm{TOT}}$, the LR solution at $r=r^+_{\mathrm{TOT}}$, as well as the nearby FPO trajectories, must be physically excluded. 
As a result, the LR solution at $r=r^-_{\mathrm{TOT}}$ and the unstable FPOs attached to it will not form an enclosed contour when transformed into the celestial coordinates $(\eta, \xi)$.
This is shown in the right plot of Fig.~\ref{fig2_half_ring}, the dashed orange curve indicates the visible section of the ring structure, while the dotted gray part is cut off since $r_{\mathrm{sh}} > r^-_{\mathrm{TOT}}$.
Moreover, even though the above incomplete arc is located in the region outside of the black hole shadow, we argue it gives rise to a nontrivial effect.
Similar to the role that the FPOs play in a horizonless compact object~\cite{agr-shadow-28}, it may lead to an infinite number of Einstein rings accumulated in the vicinity of the arc, on both the inside and outside.
The novelty for the present case is that the above structure does not form an enclosed curve, as it is truncated by the thin layer of dark matter at $(\eta(r_{\mathrm{sh}}), \xi(r_{\mathrm{sh}}))$.
It is noted that the classical Einstein ring\footnote{To be more precise, it is the primary critical curve associated with the primary caustic.} of the Kerr metric with $m=M_{\mathrm{TOT}}, a=a_{\mathrm{TOT}}$ corresponds to the outmost contour of the above structure.

\begin{figure}
\begin{tabular}{cc}
\vspace{0pt}
\begin{minipage}{225pt}
\centerline{\includegraphics[width=200pt]{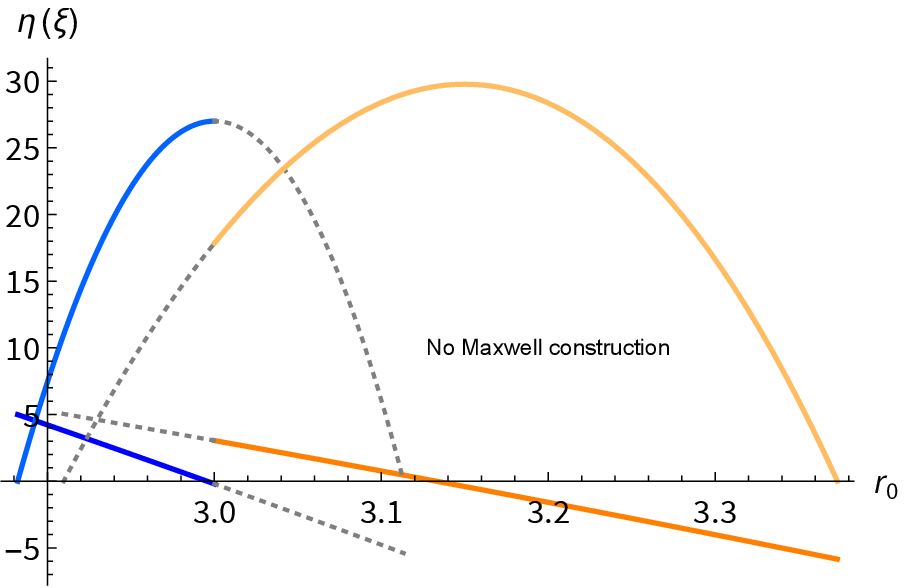}}
\end{minipage}
&
\begin{minipage}{225pt}
\centerline{\includegraphics[width=200pt]{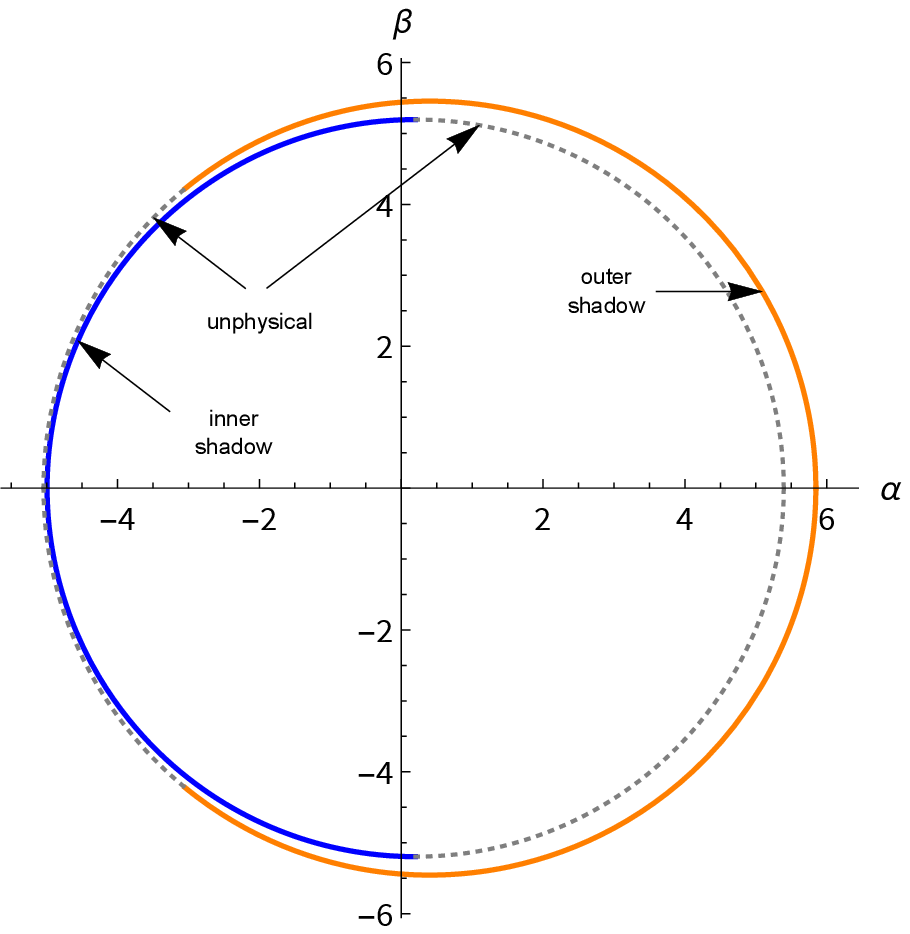}}
\end{minipage}
\end{tabular}
\renewcommand{\figurename}{Fig.}
\caption{(Color online)
The fractured black hole shadow is featured by two (mostly disconnected) sections.
We adopt the same convention for the line types and colors as Fig.~\ref{fig1_cusp}.
Left: No Maxwell construction can be established for this case.
Right: The resultant black hole shadow edge, shown in solid curves, is apparently fractured.
}
\label{fig3_fractured_shadow}
\end{figure}

Last but not least, let us discuss the scenario regarding the third set of metric parameters given in Tab.~\ref{tb_shadow}.
The calculated black shadow is presented in Fig.~\ref{fig3_fractured_shadow}.
In the present case, we note that the resultant solutions of the LR radii are entirely identical to those of Fig.~\ref{fig2_half_ring}.
However, due to the difference in the location of the thin shell, different sections of the FPO branches are truncated.
As a result, as indicated in the right plot of Fig.~\ref{fig3_fractured_shadow}, the original black hole shadow defined by the black hole sitting at the center cannot form a complete circle.
It has to be compensated by part of the unstable branch of FPOs of the outer region, namely, the Kerr metric perceived from infinity.
Since there is no explicit Maxwell construction, the two parts of the shadow arc seem to be disconnected.
This, apparently, leads to a contradiction, since the shadow edge must be a continuous curve.
We understand that in practice, the thickness of the thin layer of dark matter, though nonetheless insignificant, must be finite.
As a result, any continuous matter distribution will dictate a specific form of the shadow edge which continuously connects the endpoint of the solid blue curve to that of the solid orange curve.
Visually, the resulting black hole is featured by a sharp edge as we refer to it as ``fractured''.  

\section{Further discussions and concluding remarks} \lb{section4}

To summarize, in this work, we showed that rich features concerning the black hole shadow can be obtained using a simple but analytic toy model.
The model we devised consists of a thin shell of slowly rotating dark mass wrapping around a slowly rotating Kerr black hole while preserving the axisymmetry of the system.
It is found that the resulting metric possesses two disconnected branches of unstable FPOs.
Moreover, their interplay with the location of the dark matter layer leads to various features such as the cuspy and fractured black hole shadow edge.
In terms of the Maxwell construction, an analogy was made between the transition among different branches of FPOs and that occurs in a thermodynamic system.
In particular, we have investigated three different spacetime configurations aiming at illustrating exhaustively all the features of the present model.
The first set of parameters is designated to the case where the Maxwell construction can be established.
The parameters are particularly chosen so that both unstable branches contribute to form an enclosed shape, which subsequently defines the contour of the black hole shadow.
The point of transition corresponds to a pair of cusps on the shadow edge.
Moreover, the remaining FPOs are mapped onto the eyelash shape extension of shadow edge on the celestial coordinates, reminiscent of the metastable states in a thermodynamical system.
The other two sets of parameters are dedicated to the cases where the Maxwell construction cannot be established.
The second set leads to a scenario where the shadow, solely defined by one branch of FPOs related to the Kerr black hole sitting at the center, is enclosed by an incomplete arc of Einstein rings.
For the third set of parameters, again, both unstable branches of FPOs contribute to the shadow.
However, since there is no Maxwell construction, the two sections of the shadow edge are apparently disconnected, giving rise to a fractured shadow edge.
The above choices of metric parameters are representative of different physical outcomes implied by the proposed model.
In terms of which, we show that interesting physics can be realized in a rather straightforward framework.

A few additional comments are necessary to clarify the difference and novelty between the method proposed in the present study and those in the existing literature.
One may understand that the procedure to calculate the black hole shadow consists essentially of two parts.
On the one hand, one needs to identify the relevant FPOs, which largely furnish the boundary of the black hole shadow. 
On the other, these FPOs should be mapped onto the celestial coordinates ($\alpha$ and $\beta$) of the observer's local sky, as given by Eqs.~\eqref{alphabeta}. 
The above formalism was introduced in~\cite{book-blackhole-DeWitt, agr-shadow-02}, and later further developed by many authors~\cite{agr-shadow-03, agr-shadow-04, agr-shadow-06, agr-strong-lensing-shadow-08, agr-strong-lensing-shadow-09}.
The method proposed in the present study concerns some subtlety first part of the procedure.
On the one hand, the shadow edge may not be entirely furnished by FPOs, as shown to be the case where there is no horizon~\cite{agr-strong-lensing-shadow-08, agr-strong-lensing-shadow-09}. 
Such a special section of the shadow edge can be furnished by either principal null geodesics or particular escaping geodesics. 
On the other hand, an unstable FPO may not constitute a pixel on the shadow edge.
This was demonstrated in~\cite{agr-shadow-19}, where the cause was attributed to the emergence of a stable branch of FPOs.
We show that such a scenario can be further explored by elaborating on a thermodynamic analogy, which naturally provides a more transparent interpretation of the physical content.
For the case of the Kerr metric, an unstable FPO corresponds to an orbit that sits at the local maximum of the radial effective potential.
While such an orbit is {\it locally} favorable as to furnish the edge of the black hole shadow, its role eventually also depends on the {\it global} properties of the effective potential.
To be specific, it also has to be an escaping null geodesic in order to reach an asymptotic observer and its position on the celestial coordinates must be bounded from inside.
In this sense, the above scenario is analogous to the condition of instability and phase transition in thermodynamics.
The local stability is only a necessary condition for a thermodynamic state to be in equilibrium.
At constant temperature and volume, a more general requirement is that the free energy must be globally minimized so that the state is not subjected to any phase transition.
The condition for the states that marginally satisfy the last criterion is the well-known Gibbs conditions for the phase transition in a thermodynamic system.
In the present context, they possess the form of Eq.~\eqref{Maxwell_01} and are visually presented in Fig.~\ref{fig1_cusp} in terms of the Maxwell construction.
To summarize, we proposed a method to determine which FPO should be counted in the shadow calculation, aiming primarily at the scenarios when some FPOs are irrelevant to constitute the shadow.
In other words, our method further refines the traditional ones initiated by Bardeen and later developed by several authors, which is tailored to handle the specific cases discussed above.

It is worth noting that the spacetime configuration under consideration, and in particular, the presence of the discontinuity in the effective potential, is indeed physically relevant.
In what follows, we further elaborate on a few realistic scenarios where discontinuity plays a pertinent role.
First, discontinuity makes its appearance in dark halos.
By using the N-body numerical calculations, discontinuity, dubbed ``cusp'', was observed in the resultant halo profile in the context of $\Lambda$CDM models~\cite{agr-dark-matter-06, agr-dark-matter-07}.
Although such a feature in the dark halos was largely considered as a ``problem'', it has also been pointed out that the rotation curves of specific galaxies are largely compatible with the presence of discontinuous dark halos~\cite{agr-dark-matter-08}.
On the other hand, in the outer region of the dark matter distribution, a sudden drop in the density profile was also spotted numerically~\cite{agr-dark-matter-21, agr-dark-matter-24}.
The latter is referred to as ``splashback'' in the literature, which also gives rise to a discontinuity in the outskirt of the profile.
Intuitively, when a rotating black hole is surrounded by these types of dark halos, the corresponding shadow is subsequently subjected to the characteristics investigated in the present study.
Second, discontinuity is also a pertinent feature in the context of a dynamically collapsing setup.
In the study of the time evolution of a spherical collapsing matter, where the backreaction regarding evaporation is taken into consideration~\cite{agr-collapse-thin-shell-11,agr-collapse-thin-shell-12}, the interior metric was found to possess discontinuity.
Although the present study does not explicitly involve dynamic black hole metrics, it is plausible that the role of discontinuity essentially remains similar. 
As the third and last scenario, one might argue that discontinuity constitutes an important assembly component in the context of exotic compact objects (ECOs).
Typical examples include the gravatar~\cite{agr-eco-gravastar-02, agr-eco-gravastar-03, Alestas2020} and wormhole~\cite{agr-wormhole-10,Dias:2010uh}.
Indeed, the concept of a discontinuous thin shell is essential to construct the throat of traversable wormholes using the cut-and-paste procedure~\cite{agr-wormhole-11}, which allows one to confine exotic matter in a limited part of the spacetime.
Subsequently, the resultant metric naturally possesses a discontinuity. 
Also, as a horizonless ECO, gravastar is characterized by non-perturbative corrections to the near-horizon external geometry of the corresponding black hole metric.
In the original picture proposed by Mazur and Mottola, it is implemented by introducing different layers of matter compositions with distinct equations of state, and therefore, it naturally leads to discontinuity.
Such a construction scheme has been subsequently adopted by most generalizations of the model, inclusively for rotating metrics.
In this regard, ECOs equipped with unstable FPO have also been a topic of much interest~\cite{Sakai2014, Ohgami2017, Shaikh2018}.
Based on the above discussions, one concludes that discontinuity can be viewed as an astrophysically relevant feature in the black hole as well as ECO metrics.

In the previous discussions, we have considered a simplified scenario where a discontinuity is planted by including a thin layer of dark matter surrounding the black hole at a given radial coordinate.
To a certain extent, the proposed metric is somewhat exaggerated when compared to the cuspy dark matter halos~\cite{agr-dark-matter-06, agr-dark-matter-07}.
However, the main goal of the present study is to illustrate that some interesting features of the black hole shadow can be understood in terms of a barebone approach.
Moreover, one may argue that most of our results will remain valid when one generalizes the metric given in Eq.~\eqref{metric_Kerr}-\eqref{Ma_cut} to that regarding a more realistic matter distribution.
To be specific, one may consider a thin but continuous matter distribution is used to replace the dark matter shell with infinitesimal thickness while maintaining the axisymmetry.
For the case where the Maxwell construction can be encountered, such as that studied in Fig.~\ref{fig1_cusp}, the two endpoints of the metastable part of the FPO branches will be connected (probably by a branch of stable FPOs).
On the left plot of Fig.~\ref{fig1_cusp}, this corresponds to a curve that joins continuously between the endpoint of the dashed yellow curve and that of the dashed blue curve.
It is noted that the Maxwell construction will remain unchanged as long as the section of the metric involving the rectangle stays the same.
This is indeed the case if the matter distribution is confined inside the interval $\left[ r^{\mathrm{cusp}}_{\mathrm{BH}}, r^{\mathrm{cusp}}_{\mathrm{TOT}}\right]$.
Similar arguments can be given to the other two cases where the Maxwell construction cannot be established.
In particular, as discussed in the last section, for the scenario investigated in Fig.~\ref{fig3_fractured_shadow}, a finite thickness is required to properly evaluate the shadow edge between the two rings.
The main advantage to introduce an infinitesimally thin layer of dark matter is that the discontinuity brings mathematical simplicity, as well as a more transparent interpretation of the relevant physics content.
As discussed in the appendix, the metric proposed in the present study is, in fact, an approximation up to the first order in $a$.
The discrepancy between the induced metrics projected on the hypersurface from the exterior and interior spacetimes is of second order in the rotation parameter.
Therefore, one may heuristically argue that such a small discrepancy between the two sides of the shell can be understood as a nonvanishing but insignificant thickness.
By considering the above arguments regarding the validity of Maxwell's construction for a shell of small thickness, the approximation assumed for the metric does not undermine our conclusion.
Moreover, we note that the second equality of Eq.~\eqref{alphabeta}, and subsequently, the entire equation, is valid for any asymptotically Kerr spacetime.
As a result, the Maxwell construction utilized in the present study is valid for any axisymmetric metric which asymptotically approaches a Kerr solution.
Based on the above discussions and the astrophysical significance of the Kerr-type metrics, we argue that our findings are meaningful and potentially valid on a rather general ground.

Last but not least, we make a few comments about the relation with the empirical observations of the black hole shadow, and in particular, the image of M87* obtained recently by the EHT Collaboration~\cite{agr-shadow-EHT-L05, agr-shadow-EHT-L06}.
The present work, similar to most studies in the literature, has been carried out in the context of a given spacetime configuration, for which the black hole shadow is evaluated.
On the other hand, the inverse problem is physically pertinent from a practical viewpoint.
From the measured black hole silhouette, one is expected to extract the essential information on the underlying spacetime metric.
Such a topic has been explored by several authors~\cite{agr-strong-lensing-shadow-38, agr-strong-lensing-shadow-39, agr-strong-lensing-shadow-40, agr-strong-lensing-EHT-08}.
The main idea, as proposed by Hioki and Maeda~\cite{agr-strong-lensing-shadow-38}, is to first quantify the apparent shape and distortion of the shadow in terms of characteristic parameters, such as the radius and dent $(R_s, \delta_s)$.
Subsequently, by using an appropriate scheme, the information on the black hole, such as the spin and inclination angle $(a, i)$, can be extracted from these quantities.
In the framework of Einstein's general relativity, if one presumably considers a Kerr black hole in the vacuum, the conclusion was drawn that the spin and inclination angle can be determined with reasonable precision~\cite{agr-strong-lensing-shadow-38}.
However, one encounters a few difficulties in more general as well as realistic scenarios.
As pointed out by Bambi {\it et al.}, from the apparent shape of the black hole shadow, it is rather difficult to tell apart an astrophysical Kerr black hole from a Bardeen one~\cite{agr-strong-lensing-shadow-39}.
Furthermore, there is a strong cancellation between the effect of frame dragging and that of the spacetime quadrupole in a Kerr-like metric.
As a result, the size and shape of the shadow outline depend weakly on the spin of the black hole or the orientation of the observer~\cite{agr-strong-lensing-shadow-42}.
However, it was also pointed out that the above cancellation can be largely attributed to the no-hair theorem and, therefore, the violation of the latter might substantially modify the shadow~\cite{agr-strong-lensing-EHT-08}.
Regarding the image of M87*, at the present stage, the resolution of the data is not yet desirable for quantitative analysis of the detailed features of the shadow edge. 
To be specific, the reconstructed image was shown to be rather sensitive to the specific characteristics of the crescent structure around the black hole~\cite{agr-shadow-EHT-L05, agr-shadow-EHT-L06}, while the Einstein rings and black hole shadow cannot be inferred straightforwardly from the data.
Since the cusp feature investigated in the present study resides on the specific detail of the shadow edge, it is not yet feasible at the moment.
Nonetheless, it is worth pointing out, various studies have been performed out in an attempt to extract the black hole spin parameters using the reconstructed black hole image~\cite{agr-shadow-EHT-05, agr-shadow-EHT-06, agr-shadow-EHT-07, agr-strong-lensing-shadow-40,agr-strong-lensing-EHT-08}.
In conjunction with other observations, such as EMRI and electromagnetic spectra, it is expected to extract more precise information from the black hole candidates in the near future.
In this regard, the ongoing observational astrophysics enlightens an optimistic perspective on a variety of promising frontiers.
Therefore, it is worthwhile to explore the subject further, inclusively extend the study to more realistic scenarios.

\section*{Acknowledgments}
We are indebted to Che-Yu Chen for enlightening discussions and communications.
WLQ is thankful for the hospitality of Huazhong University of Science and Technology, where a significant portion of this work was carried out.
We gratefully acknowledge the financial support from
Funda\c{c}\~ao de Amparo \`a Pesquisa do Estado de S\~ao Paulo (FAPESP),
Funda\c{c}\~ao de Amparo \`a Pesquisa do Estado do Rio de Janeiro (FAPERJ),
Conselho Nacional de Desenvolvimento Cient\'{\i}fico e Tecnol\'ogico (CNPq),
Coordena\c{c}\~ao de Aperfei\c{c}oamento de Pessoal de N\'ivel Superior (CAPES),
and National Natural Science Foundation of China (NNSFC) under contract Nos. 11805166, 11775036, and 11675139.
A part of this work was developed under the project Institutos Nacionais de Ciências e Tecnologia - Física Nuclear e Aplicações (INCT/FNA) Proc. No. 464898/2014-5.
This research is also supported by the Center for Scientific Computing (NCC/GridUNESP) of the S\~ao Paulo State University (UNESP).

\appendix
\section{The junction condition of the thin shell}

In this appendix, we show that the metric proposed in Eqs.~\eqref{metric_Kerr} and~\eqref{Ma_cut} are physically meaningful at the slow rotation limit, namely, $\left|a\right| \ll 1$.

To be specific, one validates the Israel-Lanczos-Sen's junction conditions~\cite{agr-collapse-thin-shell-03} which deal with the case when a hypersurface $\Sigma$ partitions spacetime into two regions $\mathscr{V}^+$ and $\mathscr{V}^-$.
For such a separation to be physically meaningful, the tangential projections of metrics on $\Sigma$, namely, the induced metrics, must be the same on both sides of the hypersurface.
On the other hand, in the normal direction, the metric might be discontinuous.
The amount of discontinuity, in terms of the extrinsic curvature, gives rise to the energy momentum tensor on the hypersurface.
Furthermore, if one explicitly indicates the specific form of the equation of state, then the dynamical equation of motion of the hypersurface can be determined.

It is noted that even though the metrics in 
\begin{eqnarray}
\mathscr{V}^+=\left\{x^\mu: r>r_{\mathrm{sh}}\right\} 
\lb{V1}
\end{eqnarray}
and
\begin{eqnarray}
\mathscr{V}^-=\left\{x^\mu: r<r_{\mathrm{sh}}\right\} 
\lb{V2}
\end{eqnarray}
both satisfies the vacuum Einstein equation, for arbitrary $a$, the two induced metrics on the hypersurface $\Sigma$ defined by 
\begin{eqnarray}
\Phi(x^\mu) =0 ,
\lb{Phi1}
\end{eqnarray}
where
\begin{eqnarray}
\Phi(x^\mu)=r-r_{\mathrm{sh}}
\lb{Sigma1}
\end{eqnarray}
are ``incompatible''.
This is because the induced metrics from both sides cannot be put into isometric correspondence.
This difficulty is well-known and closely related to that explored extensively in the literature, regarding the possible source for the Kerr metric.
According to Krasiński~\cite{agr-bh-Kerr-07}, there are essentially four classes of approaches. 
The class relevant to the present scenario is the third one where one attempts to construct approximate physically acceptable configurations matched to the exterior Kerr metric.
Rotating thin shell as an approximate source of the Kerr metric was initiated by Cohen and Brill~\cite{agr-bh-Kerr-05} and extended by la Cruz and Israel~\cite{agr-bh-Kerr-06}.
Those studies indicated that metrics similar to that given in Eqs.~\eqref{metric_Kerr} and~\eqref{Ma_cut} are feasible at the slow rotating limit. 

Following this line of thought, one may generalize the above result and argue that a thin rotating shell $\Sigma$ separates two slowly rotating spacetimes $\mathscr{V}^+$ and $\mathscr{V}^-$.
To be more specific, in what follows, we show that Israel's first junction condition is indeed satisfied up to first order in $a$. 
This can be accomplished by explicitly evaluating and comparing the induced metrics for both spacetimes $\mathscr{V}^+$ and $\mathscr{V}^-$.
Here, the interior and exterior spacetimes $\mathscr{V}^\pm$ are defined in Eqs.~\eqref{V1} and~\eqref{V2} and the shell $\Sigma$ is defined by Eq.~\eqref{Sigma1}.
At the slow rotation limit, one can expand the metrics in terms of $a$ to first order~\cite{agr-bh-Kerr-06, book-general-relativity-Poisson}.
The metric of the exterior spacetime $\mathscr{V}^+$ gives
\begin{eqnarray}
ds_+^2=-f_+ dt_+^2 +g_+^{-1} dr^2+r^2 d\Omega^2 -\frac{4M_+a_+}{r}\sin^2\theta dt_+ d\varphi ,
\lb{ds_plus}
\end{eqnarray}
where
\begin{eqnarray}
f_+&=&\frac{1-\frac{2M_{\mathrm{TOT}}}{r}}{1-\frac{2M_{\mathrm{TOT}}}{r_{\mathrm{sh}}}}, \nonumber\\
g_+&=&1-\frac{2M_{\mathrm{TOT}}}{r},\nonumber\\
M_+&=&\frac{M_{\mathrm{TOT}}}{\sqrt{1-\frac{2M_{\mathrm{TOT}}}{r_{\mathrm{sh}}}}}, \nonumber\\
a_+&=&a_{\mathrm{TOT}}\nonumber,
\lb{f_plus}
\end{eqnarray}
and one has also rescaled the time coordinate by 
\begin{eqnarray}
t_+={t}{\sqrt{1-\frac{2M_{\mathrm{TOT}}}{r_{\mathrm{sh}}}}} \nonumber
\end{eqnarray}
Therefore, when viewed from the exterior, the shell's induced metric reads
\begin{eqnarray}
ds_\Sigma^2=-dt_+^2 +r_{\mathrm{sh}}^2 d\Omega^2 -\frac{4M_+a_+}{r_{\mathrm{sh}}}\sin^2\theta dt_+ d\varphi .
\lb{ds_Sigma_plus}
\end{eqnarray}

Now, we can show that the above induced metric essentially possesses spherical geometry by properly introducing the ``rotating''
\begin{eqnarray}
\psi_+=\varphi-\Omega_+ t_+ \nonumber ,
\end{eqnarray}
where
\begin{eqnarray}
\Omega_+ =\frac{2M_+a_+}{r_{\mathrm{sh}}^3} \nonumber .
\end{eqnarray}
It is readily shown, by using $y^a=(t_+,\theta,\psi_+)$ as the coordinates on the shell, the following $(2+1)$ Minkowski metric on $\Sigma$,
\begin{eqnarray}
h^+_{ab}dy^a dy^b=-dt_+^2 +r_{\mathrm{sh}}^2 (d\theta^2+\sin^2\theta d\psi_+^2) .
\lb{h_ab_plus}
\end{eqnarray}

On the other hand, by pratically identical arguments, one derives the induced metric for the interior spacetime $\mathscr{V}^-$ 
\begin{eqnarray}
h^-_{ab}dy^a dy^b=-dt_-^2 +r_{\mathrm{sh}}^2 (d\theta^2+\sin^2\theta d\psi_-^2) ,
\lb{h_ab_minus}
\end{eqnarray}
where
\begin{eqnarray}
\psi_-&=&\varphi-\Omega_- t_- \nonumber ,\\
\Omega_- &=&\frac{2M_-a_-}{r_{\mathrm{sh}}^3}, \nonumber \\
t_-&=&{t}{\sqrt{1-\frac{2M_{\mathrm{BH}}}{r_{\mathrm{sh}}}}}, \nonumber \\
M_-&=&\frac{M_{\mathrm{BH}}}{\sqrt{1-\frac{2M_{\mathrm{BH}}}{r_{\mathrm{sh}}}}}, \nonumber\\
a_-&=&a_{\mathrm{BH}}\nonumber .
\end{eqnarray}
Apparently, Eq.~\eqref{h_ab_plus} and~\eqref{h_ab_minus} are isometric, which means that the tangencial projections of the spacetimes metrics on $\Sigma$ is continuous. 
On the other hand, in the normal direction, there is a discontinuity, measured by that of the extrinsic curvature, which gives rise to the energy-momentum tensor on the shell~\cite{agr-collapse-thin-shell-03}
\begin{eqnarray}
S_{ab}=-\frac{\epsilon}{8\pi}\left(\left[K_{ab}\right]-\left[ K\right]h_{ab}\right) ,
\end{eqnarray}
where $K_{ab}=n_{\alpha;\beta}e^\alpha_ae^\beta_b$ is the extrinsic curvature, $e^\alpha_a\equiv \frac{\partial x^\alpha}{\partial y^a}$, and the normal vector $n_\alpha\partial^\alpha=-\partial^r$ with $\epsilon=-1$ for our present case.
If the equation of state of the shell is further given, its dynamic evolution is subsequently governed by the Einstein equation.
In the main text, our toy model has been contructed based on the above slow rotating case, which implies $\left|a_{\mathrm{BH}}\right|, \left|a_{\mathrm{TOT}}\right| \ll 1$.
In reality, there is some difference between the two induced metrics, whose magnitude is of the order $a^2$.
Intuitively, such a small discrepancy can be compensated by a nonvanishing but nonetheless thin shell.

\bibliographystyle{JHEP}
\bibliography{references_reply, references_qian}

\end{document}